\documentclass[12pt]{iopart}

\usepackage{graphicx}
\begin{document}

\newcommand{\ket}[1]{\left| #1 \right\rangle}
\newcommand{\bra}[1]{\left\langle #1 |\right}

\title{Adiabatic rapid passage two-photon excitation of a Rydberg atom}

\author{Elena Kuznetsova$^{\dag,\ddag}$, Gengyuan Liu$^\S$ and Svetlana A. Malinovskaya$^\S$}
\address{$^\dag$ Russian Quantum Center, Skolkovo, 143025, Russia}
\address{$^\ddag$ ITAMP, Harvard-Smithsonian Center for Astrophysics, Cambridge, MA 02138, USA}
\address{$^\S$ Department of Physics and Engineering Physics, Stevens Institute of Technology, Hoboken, NJ 07030, USA}

\ead{smalinov@stevens.edu}
\begin{abstract}
We considered the two-photon adiabatic rapid passage excitation of a single atom from the ground to a Rydberg state. Three schemes were analyzed: both pump and Stokes
fields chirped and pulsed, only the pump field is chirped, and only the pump field is pulsed and chirped while the Stokes field is continuous wave (CW). In all three cases high
transfer efficiencies $>99\%$ were achieved for the experimentally realizable Rabi frequencies and the pulse durations of the fields.

\end{abstract}

\maketitle

\section{Introduction}
Cold atoms in highly excited Rydberg states have been attracting a significant interest in recent years due to their exaggerated properties. Most notable is
their ability to experience a strong and a long-range
van der Waals or dipole-dipole interaction with the interaction strength scaling as $\sim n^{11}$ and $\sim n^{4}$, respectively, with the principal quantum number
$n$ \cite{Gallagher}. The strong interaction between Rydberg atoms was suggested as a way to realize the  two-qubit gates in quantum computing \cite{Rydb-QC} and quantum simulation \cite{Rydb-simul}
setups utilizing neutral atoms. It can be used to produce highly entangled cluster states \cite{Cluster-paper} and novel quantum many-body states such as Rydberg crystals \cite{Rydb-many-body}.
The Rydberg-Rydberg interaction also drastically modifies absorptive and dispersive properties of an atomic gas
with Rydberg excitations, leading to giant collective non-local nonlinearities exceeding those of EIT-media \cite{Rydb-nonl}.
The interaction in Rydberg states may also be used in the deterministic single photon generation \cite{Rydb-single-phot}, the atom-light
entanglement \cite{Rydberg-photon-entanglement}, a single photon-photon interaction \cite{Rydb-phot-phot}, etc.

Many of these applications involve the dipole blockade effect \cite{Rydb-QC}. The idea behind the dipole blockade is to excite a single atom to a
Rydberg state which will prevent (blockade) excitation to the Rydberg state of other atoms due to a shift of multiply excited state by the dipole-dipole interaction.
In the blockaded regime only one atom has to be excited to the Rydberg state. Rydberg states are
the states with large principal quantum numbers $n\ge 30$. For alkali metal atoms, typically used in
Rydberg experiments, the direct transition from the ground S$_{1/2}$ to the Rydberg state lies in the
ultraviolet range, where powerful lasers are not readily available. This limitation was overcome using a two-photon excitation
via far-detuned intermediate P$_{1/2,3/2}$ states. In this case the first step is the excitation of the D$_{1}$ or
D$_{2}$ transition with a wavelength in the range 600-1000 nm, while the second step is the excitation from  the P$_{1/2,3/2}$ to the Rydberg state by the wavelength in the 300-500 nm range.

An atom can be transferred to the Rydberg state with a two-photon $\pi$ pulse, which requires precise adjustment
of the pulse Rabi frequency and the duration \cite{Rydberg-pi-pulse1,Rydberg-pi-pulse2}. A more robust approach is to excite the atom using a chirped
pulse via adiabatic rapid passage (ARP) \cite{ARP1}. In ARP the precise adjustment of the pulse
Rabi frequency and the duration are not required, they just have to satisfy the adiabaticity condition
$|d\Delta /dt| \ll \Omega^{2}$, where $\Delta$ is the detuning from the Rydberg state, and $\Omega$ is the Rabi frequency. A chirped pulse one-photon excitation of a
single atom to a Rydberg state was discussed in \cite{Ryabtsev1,Ryabtsev2}, where it was
shown that ARP can provide high transfer efficiency and is robust to small variations of the pulse duration and
the Rabi frequency contrary to the $\pi$ pulse. In the case when a single Rydberg
atom has to be excited out of an ensemble of $N$ atoms, ARP has a crucial advantage over the $\pi$ pulse. The $\pi$ pulse excites the ensemble from a collective ground state to a symmetric collective excited state
with one atom in the Rydberg and all other
atoms in the ground state. The Rabi frequency for such a transition is collectively enhanced by $\sqrt{N}$: $\Omega_{N}=\Omega_{1}\sqrt{N}$, where $\Omega_{1}$ is the single
atom Rabi frequency. Typically the exact number $N$ of atoms in the ensemble is not known, which introduces an error in $\Omega_{N}$. It was pointed out in
\cite{Ryabtsev1} that ARP provides an efficient excitation to the collective state even when the atom number is not known.

The one-photon ARP can provide an efficient transfer to a Rydberg state, however, experimentally it may be inconveniently to do because it would require a chirped pulse
in the ultraviolet range. As a more robust approach we consider implementing the two-photon chirped pulse excitation scheme. In this work, we analyze a
possibility of the two-photon chirped pulse excitation of a single atom to a Rydberg state by using the pump and Stokes fields following three schemes, when both the pump and Stokes pulses are chirped, then,  only pump pulse is chirped, and finally, the pump pulse is chirped and the Stokes is CW.

\section{ARP in a three-level ladder system: theory}

We analyze a three-level ladder atomic system, shown in Fig.\,\ref{fig:Ladder-scheme}a. It interacts with two laser fields: the pump field which is close to the resonance with the
$\ket{g}-\ket{i}$ transition, and the Stokes field, close to the resonance with the $\ket{i}-\ket{r}$ transition. We assume that the fields are detuned from the fast decaying
intermediate state $\ket{i}$ to minimize its population. We also assume that the pump (and Stokes) field may be chirped in a general case, such that the fields' electric components are given by
\begin{eqnarray}
{\cal E}_{\rm p}=\frac{E_{\rm p}(t)}{2}\rme^{-i(\omega_{\rm p}t+\alpha t^{2}/2)}+c.c., \,
{\cal E}_{\rm S}=\frac{E_{\rm S}(t)}{2}\rme^{-i(\omega_{\rm S}t+\beta t^{2}/2)}+c.c.,
\end{eqnarray}
where $\omega_{\rm p(\rm S)}$ is the frequency of the pump (Stokes) field, $E_{p(S)}(t)$ is the field amplitude, which may be time-dependent, and $\alpha$ and $\beta$ are
the chirp
rates of the pump and Stokes fields. Neglecting decay from the excited states and aiming to illustrate the physical mechanism of the population transfer we write the Schr$\ddot{o}$dinger equation
$i\hbar \partial \ket{\Psi}/\partial t=H\ket{\Psi}$ for the wavefunction $\ket{\Psi}=\sum_{k=\rm g,i,r}C_{k}\ket{k}$, where the Hamiltonian is $H=H_{0}+H_{\rm int}$
with $H_{0}=\hbar\omega_{\rm ig} |i><i| + \hbar\omega_{\rm rg}|r><r|$ and $H_{\rm int}=-\vec{\mu}_{\rm ig}\vec{E}_{\rm p}|i><g|-\vec{\mu}_{\rm ri}\vec{E}_{\rm S}|r><i|+H.c$.
 Setting the time-dependence of the state amplitudes as
$C_{\rm g}=c_{\rm g}$, $C_{\rm i}=c_{\rm i}\exp{(-i(\omega_{\rm p}t+\alpha t^{2}/2))}$, $C_{\rm r}=c_{\rm r}\exp{(-i(\omega_{\rm p}t+\omega_{\rm S}t+\alpha t^{2}/2+\beta t^{2}/2))}$
we get a set of equations
\begin{eqnarray}
\rmi \frac{\rmd c_{\rm g}}{\rmd t}=-\frac{\Omega_{\rm p}^{*}(t)}{2}c_{\rm i}, \nonumber \\
\rmi \frac{\rmd c_{\rm i}}{\rmd t}=\Delta(t)c_{\rm i}-\frac{\Omega_{\rm p}(t)}{2}c_{\rm g}-\frac{\Omega_{\rm S}^{*}(t)}{2}c_{\rm r}, \nonumber \\
\rmi \frac{\rmd c_{\rm r}}{\rmd t}=\delta(t)c_{\rm r}-\frac{\Omega_{\rm S}(t)}{2}c_{\rm i},
\end{eqnarray}
where we denoted one- and two-photon detunings as $\omega_{\rm ig}-\omega_{\rm p}-\alpha t=\Delta(t)$, $\omega_{\rm rg}-\omega_{\rm p}-\omega_{\rm S}-\alpha t -\beta t=\delta(t)$ (see Fig.\,\ref{fig:Ladder-scheme}a),
and the pump and Stokes Rabi frequencies as $\Omega_{\rm p}=E_{\rm p}\mu_{\rm ig}/\hbar$, $\Omega_{\rm S}=E_{\rm S}\mu_{\rm ri}/\hbar$.

Next we assume that the one-photon detuning is sufficiently large, $\Delta \gg \Omega_{\rm p}, \Omega_{\rm S}$, which allows us to eliminate the intermediate state $\ket{i}$
by setting $c_{\rm i}\approx (\Omega_{\rm p}c_{\rm g}+\Omega_{\rm S}^{*}c_{\rm r})/2\Delta$. After the elimination of $c_{i}$ we arrive at two equations for the probability amplitudes of
the ground and the Rydberg states:
\begin{eqnarray}
\label{eq:Schr-eq}
\rmi \frac{\rmd c_{\rm g}}{\rmd t}=-\Delta_{g}c_{\rm g}-\tilde{\Omega}^{*}c_{\rm r}, \nonumber \\
\rmi \frac{\rmd c_{\rm r}}{\rmd t}=\left(\delta -\Delta_{r}\right)c_{\rm r}-\tilde{\Omega}c_{\rm g},
\end{eqnarray}
where we introduced the Stark shifts of the ground $\Delta_{g}=|\Omega_{p}|^{2}/4\Delta$ and the Rydberg $\Delta_{r}=|\Omega_{S}|^{2}/4\Delta$ states, and the effective two-photon
Rabi frequency $\tilde{\Omega}=\Omega_{\rm p}\Omega_{\rm S}/2\Delta$.
We find the atom-field dressed states and their energies by setting $\rmi \left(\begin{array}{c} \rmd c_{\rm g}/\rmd t \\ \rmd c_{\rm r}/\rmd t \end{array} \right)=\lambda \left(\begin{array}{c} c_{\rm g} \\ c_{\rm r} \end{array} \right)$.
 The dressed state energies read
\begin{eqnarray}
\lambda_{\pm}=-\Delta_{g}+\frac{\delta-\Delta_{r}+\Delta_{g}}{2}
\pm \left(\left(\frac{\delta-\Delta_{r}+\Delta_{g}}{2}\right)^{2}+\left(\frac{\tilde{\Omega}}{2}\right)^{2}\right)^{1/2}. \nonumber
\end{eqnarray}

We denote the effective two-photon detuning taking into account the Stark shifts of the ground and Rydberg states as
$\tilde{\Delta}=\delta-(|\Omega_{\rm S}|^{2}-|\Omega_{\rm p}|^{2})/4\Delta$.
We also assume that $\Omega_{\rm p}$, $\Omega_{\rm S}$ are real, and
introduce the rotation angle $\theta(t)$, describing the evolution of the dressed states in time, given by
$\tan \theta =\left(\left(\tilde{\Omega}^{2}+\tilde{\Delta}^{2}\right)^{1/2}-\tilde{\Delta}\right)/\tilde{\Omega}$. In terms of $\theta$, the dressed states may be written as follows
$\left(\begin{array}{c} \ket{+} \\ \ket{-} \end{array} \right) = \left(\begin{array}{cc} \sin \theta & -\cos \theta \\
\cos \theta & \sin \theta \end{array} \right) \left(\begin{array}{c} \ket{g} \\ \ket{r} \end{array} \right)$.
The respective dressed states have the following coefficients:
\begin{eqnarray}
&\hspace{-2cm} c_{\rm g\,\pm}=\frac{\tilde{\Omega}}{\left(\tilde{\Omega}^{2}+\left(\tilde{\Delta} \pm \left(\tilde{\Omega}^{2}+\tilde{\Delta}^{2}\right)^{1/2}\right)^{2}\right)^{1/2}}, \,
c_{\rm r\,\pm}=\mp \frac{\tilde{\Delta}\pm \left(\tilde{\Omega}^{2}+\tilde{\Delta}^{2}\right)^{1/2}}{\left(\tilde{\Omega}^{2}+\left(\tilde{\Delta} \pm \left(\tilde{\Omega}^{2}+\tilde{\Delta}^{2}\right)^{1/2}\right)^{2}\right)^{1/2}}. \nonumber
\end{eqnarray}

Next, we analyze two limits:
1) $\tilde{\Delta}>0$, i.e. the fields are red detuned from the two-photon resonance, and $\tilde{\Delta} \gg \tilde{\Omega}$. In this limit,
$\tan \theta \approx \tilde{\Omega}/2\tilde{\Delta} \ll 1$, which means that $\theta\simeq 0$ and $\ket{+}\approx -\ket{r}$, $\ket{-}\approx \ket{g}$;
2) $\tilde{\Delta} <0$, i.e. the fields are blue detuned from the two-photon resonance, and $|\tilde{\Delta}| \gg \tilde{\Omega}$. In this limit,
$\tan \theta \approx 2|\tilde{\Delta}|/\tilde{\Omega} \gg 1$, and $\theta \approx \pi/2$, resulting in $\ket{+}\approx \ket{g}$, $\ket{-}\approx \ket{r}$.
This analysis shows that the system may be adiabatically transferred from the ground to the excited state via the $\ket{-}$ dressed state if the effective two-photon detuning is
negatively chirped, $\dot{\tilde{\Delta}}<0$, and via the $\ket{+}$ dressed state if  the effective two-photon detuning is positively chirped, $\dot{\tilde{\Delta}}>0$.

\section{ARP in a three-level ladder system: numerical analysis}

We numerically investigate ARP with the two-photon chirped pulse excitation of a single atom from the ground to the Rydberg state. We analyze the full atomic
evolution, including the intermediate state, using the
density matrix equation $\rmd \rho/\rmd t=(\rmi /\hbar)[\rho,H]+{\cal L}\rho$, and taking into account the population and coherence decays, described by the Lindblad term
${\cal L}\rho=\sum_{kl=\rm ir,gi}\Gamma_{kl}/2\left(\sigma_{kl}\rho\sigma_{kl}^{+}-\sigma_{kl}^{+}\sigma_{kl}\rho-\rho \sigma_{kl}^{+}\sigma_{kl}\right)$,
where $\sigma_{kl}=|k><l|$ and $\sigma_{kl}^{+}=|l><k|$ are the atomic lowering and raising operators. We assume that transitions are radiatively
broadened with coherence decay rates $\gamma_{ij}=\sum_{k}(\Gamma_{ik}+\Gamma_{jk})/2$. We analyze three cases: 1) both the Stokes and the pump fields are chirped
and pulsed; 2) both fields are pulsed, but only the pump field is chirped; 3) only the pump field is pulsed and chirped, while the Stokes field is CW.

\begin{figure}[h]
\center{
\includegraphics[width=4.5in]{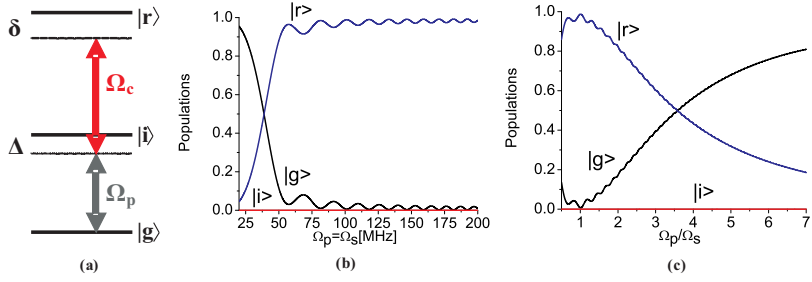}
\caption{\label{fig:Ladder-scheme} (a) Three-level ladder atomic system with a Stokes and a pump field
interacting with the $\ket{i}-\ket{g}$ and $\ket{r}-\ket{i}$ transitions, respectively. The one- and two-photon detunings are $\Delta(t)$ and
$\delta(t)$; (b) Populations of the atomic states $\ket{g}$ (black curve), $\ket{i}$ (red curve) and $\ket{r}$ (blue curve) depending on the
Rabi frequencies of the fields. Pulse durations $\tau_{p}=\tau_{S}=1$ $\mu$s, chirp rates $\alpha=\beta=4.2$ MHz/$\mu$s, one-photon detuning $\Delta=1.5$ GHz; (c) Populations of the atomic states depending on the ratio of the pump and Stokes Rabi frequencies.}
}
\end{figure}

\begin{figure}[h]
\center{
\includegraphics[width=4.in]{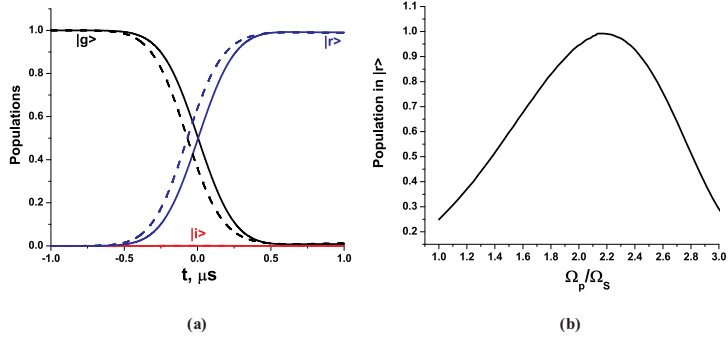}
\caption{\label{fig:case2-pop} (a) Population transfer in the case of both pump and Stokes fields pulsed and only pump field is chirped (solid curves).
Parameters of the fileds: $\Omega_{\rm p}^{0}=\Omega_{\rm S}^{0}=25$ MHz, $\tau_{\rm p}=\tau_{\rm S}=0.45$ $\mu$s, $t_{\rm p}=t_{\rm S}=0$, $t_{0}=0$,
$\Delta^{0}=\omega_{\rm ig}-\omega_{p}=1.5$ GHz, $\alpha=2$ MHz/$\mu$s. In calculations we assumed the $^{87}$Rb atom with $\ket{i}=\ket{5P_{3/2}}$ having the
population decay rate $6$ MHz and $\ket{r}=\ket{97d_{5/2}}$ having the population decay rate $\approx 3\cdot 10^{-3}$ MHz; Population transfer in the case of only pump field
pulsed and chirped and the Stokes field CW (dashed curves). Parameters of the fields: $\Omega_{\rm p}^{0}=35$ MHz, $\Omega_{\rm S}=17$ MHz, $\tau_{\rm p}=0.34$ $\mu$s, $t_{\rm p}=0$,
$t_{0}=-0.26$ $\mu$s, $\Delta^{0}=\omega_{\rm ig}-\omega_{p}=1.5$ GHz, $\alpha=2$ MHz/$\mu$s; (b) Population transfer efficiency in the latter case, when the Stokes field is CW,
depending on the ratio of the pump and Stokes fields Rabi frequencies.
}
}
\end{figure}

We consider the Gaussian time dependence of the fields $\Omega_{\rm p}=\Omega_{\rm p}^{0}\exp(-(t-t_{\rm p})^{2}/2\tau_{\rm p}^{2})$, $\Omega_{\rm S}=\Omega_{\rm S}^{0}\exp(-(t-t_{\rm S})^{2}/2\tau_{\rm S}^{2})$,
$\omega_{\rm p}=\omega_{\rm p}^{0}+\alpha(t-t_{0})$ and $\omega_{\rm c}=\omega_{\rm c}^{0}+\beta(t-t_{0})$. Figure \ref{fig:Ladder-scheme}b displays
populations of the three atomic states in the first case, when both fields are pulsed and chirped, depending on the pump and Stokes field Rabi frequency, assumed equal,
and shows that high $\sim 100\%$ transfer efficiency
from the ground to the Rydberg state may be realized. Figure \ref{fig:Ladder-scheme}c shows populations at the end of the transfer process depending
on the ratio of the pump and Stokes Rabi frequencies. One can see that the transfer efficiency is high in a wide range of Rabi frequency ratios $0.5<\Omega_{p}/\Omega_{S}<1.5$.
Small oscillations are observed due to a small nonadiabatic coupling of the dressed states in the very beginning and at the end of the pulse duration. High transfer efficiency
is realized when adiabaticity requirements $\alpha \tau_{p(S)}^{2}\gg 1$ and $\alpha << {\tilde \Omega}^{2}$ are met, which takes place for $\Omega_{p}=\Omega_{S}>50$ MHz. A similarly efficient
population transfer is achieved in the second case, when only the pump field is chirped and both pump and Stokes fields are pulsed,
shown in Figure \ref{fig:case2-pop}a by a solid curve. The dashed curve in Fig.\ref{fig:case2-pop}a shows the population transfer resulting from  the pump field pulsed and chirped, and the Stokes
field being continuous wave. High population transfer to the Rydberg state is achieved by carefully chosen field amplitudes. 
The pump and Stokes field Rabi frequencies have to be tuned close to specific values, but high transfer efficiency can be realized when their ratio is varied in the
range $1.95 < \Omega_{p}/\Omega_{S} <2.5$.

In summary, we have demonstrated a possibility of full population transfer to the Rydberg state in ultracold Rb by two-photon excitation using microsecond, chirped laser pulses and
CW laser fields. We showed that implementing linearly chirped pulses in three different combinations may lead to adiabatic passage from the ground to the Rydberg state without
populating the fast decaying intermediate state. The adiabaticity condition $|\alpha|<< \tilde{\Omega}^2$ according to the Landau-Zener formula is satisfied by choosing the
chirp rate on the order of 1 MHz/$\mu s$ and the peak Rabi frequency on the order of tens of MHz. The technique may find applications in two-atomic systems aiming to realize CNOT or two-qubit phase gate operations.

\ack{The authors gratefully acknowledge financial support from the National Science Foundation under Grant No. NSF PHY12-05454 and the Russian Quantum Center.}


\section*{References}

\begin{thebibliography}{11}

\bibitem{Gallagher} T.F. Gallagher, {\it Rydberg atoms} (Cambridge University Press, Cambridge, UK, 1994).

\bibitem{Rydb-QC} D. Jaksch, J. I. Cirac, P. Zoller, S. L. Rolston, R. Cote, M. D. Lukin, Phys. Rev. Lett. {\bf 85}, 2208 (2000);
M. D. Lukin, M. Fleischhauer, R. Cote, L. M. Duan, D. Jaksch, J. I. Cirac, P. Zoller, Phys. Rev. Lett. {\bf 87}, 037901 (2001).

\bibitem{Rydb-simul} H. Weimer, M. Muller, I. Lesanovsky, P. Zoller, H. P. Buchler, Nat. Phys. {\bf 6}, 382 (2010).

\bibitem{Cluster-paper} E. Kuznetsova, T. Bragdon, R. Cote, S. F. Yelin, Phys. Rev. A {\bf 85}, 012328 (2012).

\bibitem{Rydb-many-body} P. Schauss, M. Cheneau, M. Endres, T. Fukuhara, S. Hild, A. Omran, T. Pohl, C. Gross, S. Kuhr, I. Bloch, Nature {\bf 491}, 87 (2012).

\bibitem{Rydb-nonl} J. D. Pritchard, K. J. Weatherill, C. S. Adams, "Non-linear optics using cold Rydberg atoms", arxiv: 1205.4890.

\bibitem{Rydb-single-phot} M. Saffman, T. G. Walker, Phys. Rev. A {\bf 66}, 065403 (2002); Y. O. Dudin, A. Kuzmich, Science {\bf 336}, 887 (2012).

\bibitem{Rydberg-photon-entanglement} L. Li, Y. O. Dudin, A. Kuzmich, Nature {\bf 498}, 466 (2013).

\bibitem{Rydb-phot-phot} T. Peyronel, O. Firstenberg, Q.-Y. Liang, S. Hofferberth, A. V. Gorshkov, T. Pohl, M. D. Lukin, V. Vuletic, Nature {\bf 488}, 57 (2012).

\bibitem{Rydberg-pi-pulse1} Y. Miroshnychenko, A. Gaetan, C. Evellin, P. Grangier, D. Comparat, P. Pillet, T. Wilk, A. Browaeys, Phys. Rev. A {\bf 82}, 013405 (2010).

\bibitem{Rydberg-pi-pulse2} X. L. Zhang, L. Isenhower, A. T. Gill, T. G. Walker, M. Saffman, Phys. Rev. A {\bf 82}, 030306 (2010).

\bibitem{ARP1} S. Malinovskaya, Int. J. Quant. Chem. {\bf 107}, 3151 (2007).

\bibitem{Ryabtsev1} I. I. Beterov, D. B. Tretyakov, V. M. Entin, E. A. Yakshina, I. I. Ryabtsev, C. MacCormick, S. Bergamini, Phys. Rev. A {\bf 84}, 023413 (2011).

\bibitem{Ryabtsev2} I. I. Beterov, M. Saffman, E. A. Yakshina, V. P. Zhukov, D. B. Tretyakov, V. M. Entin, I. I. Ryabtsev, C. W. Mansell, C. MacCormick,
S. Bergamini, M. P. Fedoruk, arxiv:1212.1138.


\end{thebibliography}
\end{document}